# A multiwavelength picture of the AFGL 5142 star-forming region

Todd R. Hunter[1], Leonardo Testi[2], Gregory B. Taylor[1], Gianni Tofani[3], Marcello Felli[3], and Thomas G. Phillips[1]

[1] California Institute of Technology, Department of Physics, Math and Astronomy, 320-47, Pasadena, CA 91125, U.S.A.
[2] Dipartimento di Astronomia e Scienza dello Spazio, Università di Firenze, Largo E. Fermi 5, I-50125 Firenze, Italy
[3] Arcetri Astrophysical Observatory, Largo E. Fermi 5, I-50125 Firenze, Italy

**Abstract.** We present molecular line, $H_2O$ maser, radio continuum and near infrared maps of the bipolar outflow source AFGL 5142. The high resolution of our molecular CO observations enables us to define the morphology of the large-scale bipolar outflow into a two lobe structure extending for $\sim 2'$ on each side of the center. In the perpendicular direction, we find consistent evidence for a second, more compact ($< 0.5'$) outflow in the form of a spatial velocity offset in the CO map and of an $H_2$ jet-like structure derived from a near infrared narrow band image. On a smaller scale size, the radio and infrared continuum observations reveal the engines of the molecular outflows. The maser emission occurs near the position of the most embedded source of the cluster, IRS1. This is located at the center of the compact outflow and jets of shocked $H_2$ and coincides with an ultra compact (UC) radio continuum source (most probably an ionized stellar wind). The $H_2O$ cluster is composed of five spatial components: two are within 0.2–0.3″ from the YSO (a few hundred AU) and three are at larger distance (1.5–2″, a few thousand AU). A marginal detection of proper motion of the two more distant masers may suggest a high expansion velocity at a distance $4\,10^3$ AU from the YSO, similar to what is found in Orion KL and W49N. The brightest NIR source of the cluster (IRS2) is associated with an IRAS point source and lies along the axis of the large-scale bipolar outflow. We propose that the masers and the compact molecular outflow are powered concurrently by the wind from the YSO associated with IRS1, while the large scale outflow could be the remnant from the formation of IRS2.

**Key words:** ISM: jets and outflows – ISM: individual (AFGL 5142) – masers – stars: formation

## 1. Introduction

$H_2O$ masers and CO molecular outflows have been known to be associated in star forming regions for nearly two decades (e.g. Genzel & Downes 1977; Lada 1985). Even though these phenomena occur on two widely different physical scales ($\sim .01$ pc and $\sim 1$ pc, respectively), recent low resolution surveys have confirmed their frequent association (both spatial and kinematic) and suggested that the two are powered by a single luminous YSO (Felli et al. 1992). In a survey by Henning et al. (1992), $H_2O$ emission was found to be present toward 80% of the observed sources with known molecular outflows. The fact that these phenomena tend to appear together when observed at low resolution motivates a higher resolution search for their detailed physical connection. Presumably, the relationship between CO outflows and $H_2O$ masers is governed by the sequence of events that lead to the formation of a new star. Recent evidence indicates that the bipolar molecular outflow begins during the infall phase of protostellar evolution (see Shu et al. 1991 and references therein) and the $H_2O$ masers are very good tracers of very young stellar objects, deeply embedded in high density molecular clumps. At the same time, recent interferometric observations have suggested that $H_2O$ masers lie in much smaller scale disk-like structures around young stellar objects such as Orion KL (Matveyenko et al. 1993) and Orion IRc2 (Wright et al. 1990). In addition, proper motion studies of the W49N masers reveal outflowing motion from a common center (Gwinn et al. 1992). Whether these structures and/or motions are a common feature of $H_2O$ maser sources and how they relate to the molecular outflows are key questions in understanding the processes that occur during the formation of massive stars and can be answered only with improved resolution studies.

To explore the spatial correlation of molecular outflows with maser phenomena and to search for the earliest phases of stellar formation we follow a multiwave-length strategy (Hunter et al. 1995). We perform CO and CS mapping of massive star-forming regions with known $H_2O$ maser emission and broad CO lines. Maps with $\leq 30''$ resolution of the molecular outflows are being obtained in order to find the accurate position of the dynamical center and the energetics of the outflow. These maps are then compared to radio interferometric observations ($\theta_{beam} \sim 0.1''$) which resolve and image the $H_2O$ maser spots and, in a few cases, reveal the presence of UC radio components (Tofani et al. 1995), i.e. the unmistakable indication of a YSO. Finally, to search for direct evidence of these young stars and to obtain information on their spectral energy distributions, in cases where UC components are found, we obtain broad band and narrow band near-infrared images.

Here we present new results on the AFGL 5142 star-forming region (distance $\sim$1.8 kpc, Snell et al. 1988). The IRAS source has a bolometric luminosity of $3.8 \times 10^3 L_\odot$ (Carpenter, Snell & Schloerb 1990). A CO outflow associated with the IRAS source was found with moderate resolution observations by Snell et al. (1988). $H_2O$ maser and $NH_3$ emission were first detected toward this source by Verdes-Montenegro et al. (1989) and OH maser emission was detected by Braz et al. (1990). Using the VLA in C-configuration, Torrelles et al. (1992) resolved three $H_2O$ spots (seven velocity components) nearly coincident with a weak (0.69 mJy peak flux density) 8.4 GHz continuum source with a cometary morphology. Its radio continuum flux density is consistent with a B2 ZAMS star, assuming optically-thin free-free emission. The continuum source was also detected at 5 GHz using the VLA D-configuration with a flux density between 0.5 and 1 mJy (McCutcheon et al. 1991 and Carpenter et al. 1990). The masers and the UC source lie approximately $30''$ to the east of the IRAS position, just outside of the error box given in the Point Source Catalog, so the association between the two has been questioned (Torrelles et al. 1992). $NH_3$ observations of Estalella et al. (1993) with a $40''$ beam show evidence for a dense hot core of gas centered on the $H_2O$ maser and the weak radio continuum source, but removed from the IRAS position. We present new observations of the molecular gas, radio continuum, $H_2O$ maser and near-infrared emission that answer several questions concerning this source.

## 2. OBSERVATIONS

### 2.1. Caltech Submillimeter Observatory

The initial CSO[1] observations were performed on 14 September 1993. The source was mapped in the CO $J$=2→1 transition with the 1024 channel 50 MHz bandwidth acousto-optical spectrometer (AOS) as the backend. The CSO beamsize at this frequency is $31''$ with a main beam efficiency of 0.76. Two coverages of three fully-sampled 7 × 15-point maps ($15''$ grid) were obtained in the on-the-fly mapping mode with the telescope driven at $3''$ s$^{-1}$. The partially-overlapping maps were subsequently summed and combined to form a 17 × 15-point image with at least 10 seconds of total on-source time per point. The offset position ($30'$ north) was verified to be free of emission to a level $< 0.05$ K, which is well below the noise level of the maps. The central position of the map corresponds to the VLA $H_2O$ maser position. Five minute spectra of the $^{13}$CO $J$=2→1 emission were also taken with a $33''$ beam toward the central position and toward the peak positions in the outflow lobes. A followup on-the-fly map of the central region was taken on 2 February 1994 in the CO $J$=3→2 and CS $J$=7→6 lines simultaneously, using the 500 MHz AOS with a beamsize of $20''$ and a main beam efficiency of 0.67. Finally, a $^{13}$CO $J$=6→5 spectrum (661.068 GHz) was obtained toward the central position using the new 650 GHz SIS receiver (Kooi et al. 1994) with an $11''$ beam and a main beam efficiency of 0.30. All spectra were analyzed with the "CLASS" software package and are presented on a main beam brightness temperature ($T_{\rm mb}$) scale. We estimate our calibration to be accurate within 20% and our pointing to within $4''$.

### 2.2. VLA

The VLA[2] observations took place on 1992 November 24 in the A-configuration as part of a larger survey of $H_2O$ masers associated with CO outflows (Tofani et al. 1995). The source was observed for 7 minutes in line mode 2AD at 22 GHz, and for 5 minutes in continuum at 8.4 GHz. The velocity resolution is 0.66 km s$^{-1}$ and the observations with the A IF cover the velocity range -19 to +17 km s$^{-1}$. The D IF covers a 25 MHz bandwidth, offset 25 MHz from the line emission in order to search for 22 GHz continuum emission simultaneously with the $H_2O$ line observations. We obtained a synthesized beam of $0.12'' \times 0.11''$ (p.a.$= -55°$) and $0.28'' \times 0.27''$ (p.a.$= -66°$) at 22 and 8.4 GHz respectively.

### 2.3. NIR observations and data reduction

The near–infrared observations were carried out on 12 February 1994 (broad band) and on 17 February 1994 (narrow band), using the Arcetri NIR camera (ARNICA) mounted at the 1.5 meter infrared telescope TIRGO[3]. ARNICA is equipped with a NICMOS3 $256 \times 256$ HgCdTe panoramic detector with a scale of $\sim 0.955$ arcsec/pixel

---

[1] The Caltech Submillimeter Observatory is funded by the National Science Foundation under contract AST-9313929.

[2] The National Radio Astronomy Observatory is operated by Associated Universities, Inc., under cooperative agreement with the National Science Foundation.

[3] The TIRGO telescope is operated by the C.A.I.S.M.I.–C.N.R. on behalf of the Ministero dell'Università e della Ricerca Scientifica e Tecnologica

(Lisi et al. 1993 and Hunt et al. 1994a). All the data reduction and analysis were performed using the ARNICA (Hunt et al. 1994c) and IRAF software packages. Absolute position calibration was achieved using the coordinates of a number of stars from the Hubble Space Telescope Guide Star Catalogue that were contained in our mosaics. Due to the least square technique used to solve for the plate constants we estimate an astrometric accuracy of the order or better than one arcsecond (Testi 1993). The full width at half maximum of the Point Spread Function obtained in these observations is $\sim 2$–$3''$.

### 2.3.1. Broad band imaging

Using the three standard J, H and K broad band NIR filters we observed the target source using a "dithering" technique: in each band we take images slightly changing the telescope position, in such a way that the object of interest is inside all frames. The flat field image is obtained combining a set of frames using a median algorithm. The precise data acquisition and reduction procedure is described in Hunt et al. (1994b,c). After reduction the images are registered and combined in order to form an extended map of the observed region. The final maps are approximately $7' \times 7'$ wide. The signal to noise ratio is not constant over the mosaic, because of the observing technique, being higher at the center (covered by all the frames) and poorer at the edges. Photometric calibration was acheived observing a set of UKIRT's faint standard stars. We estimate a photometric accuracy of about 8%. Accurate photometry was made on all the detected point sources using the DAOPHOT routines in IRAF. The limiting magnitudes obtained at the edge of the mosaic were 17.3, 16.2 and 16.0 in J, H and K respectively.

### 2.3.2. Narrow band imaging

The narrow band images were taken using narrow band interference filters with a spectral resolution of $\frac{\lambda}{\Delta\lambda} \sim 100$, centered on the rest frequencies of the Brackett $\gamma$ and $H_2$ S(1) $1 \to 0$ lines (2.166 $\mu$m and 2.122 $\mu$m).

The observing technique is similar to that used for broad band imaging, except for the fact that the "dithering path" has shorter offsets so the useful field of view is much smaller in the narrow bands ($2' \times 2'$). The reduction technique was the same as that used for the broad band observations. The mosaics obtained in this case are smaller than those in the broad bands, covering $3' \times 3'$.

The continuum emission at the frequencies of the two narrow bands were estimated using the broad band filters. A synthetic continuum image was built by calculating at each position the power law spectrum which was best fit to the broad band data. The flux calibration was made assuming that a set of stars present in the field do not have line emission or absorption, and hence their flux in the narrow band image should be the same as in the synthetic continuum. In order to match the Point Spread Function of the continuum image, the narrow band images were convolved with a Gaussian function of proper width, before subtraction. Hence the narrow band images have been scaled with a proper factor to convert counts to flux units, then convolved and continuum subtracted.

## 3. RESULTS

### 3.1. Bipolar outflows

The CO $J=2\to1$ and CS $J=7\to6$ spectra of the central position ($\alpha = 05^h27^m30\overset{s}{.}0$, $\delta = 33°45'40\overset{''}{.}0$) are shown in Fig. 1. High velocity emission is apparent in both of these molecular tracers. The CO transition suffers absorption near the LSR velocity ($-3.5$ km s$^{-1}$) with a greater amount occuring toward redshifted velocities. Since no absorption is evident in the CS spectrum, the CO absorption is probably occuring in the cooler and less dense foreground gas.

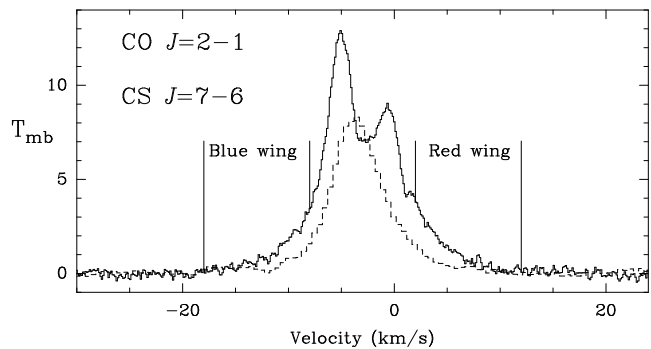

**Fig. 1.** The solid line is the CO $J=2\to1$ spectrum and the dashed line is the CS $J=7\to6$ spectrum toward the $H_2O$ maser position. The blueshifted (-18 to -8 km s$^{-1}$) and redshifted (2 to 12 km s$^{-1}$) velocity ranges correspond to the CO outflow map in Fig. 2.

The high-velocity CO outflow map is presented in the upper panel of Fig. 2 overlayed on the K–band mosaic with the blueshifted wing (-18 to -8 km s$^{-1}$) shown in solid contours and the redshifted wing (2 to 12 km s$^{-1}$) shown in dashed contours.

In this map, each lobe has a local emission peak near the outflow origin and near the outer end. In addition, the blueshifted and redshifted emission peaks near the origin are separated by 14$\pm$2$''$ along a position angle +44° (i.e. approximately perpendicular to the axis of the extended outflow). To explore the velocity structure in more detail, channel maps of the CO emission are shown in Fig. 3. The channel maps reveal a more compact outflow structure oriented almost perpendicular to the extended outflow with the centers of the two components displaced by a few arcseconds. Also plotted in Fig. 3 are the suggested axes of the two outflows (position angles: +34° and −47°) and the positions of two peculiar NIR sources that will be discussed in section 3.4. High velocity $^{13}$CO emission was not detected at the $^{12}$CO peak positions in the outflow lobes, even when averaged over the entire wing to an integrated noise level of 75 mK km s$^{-1}$. Since this corresponds to a factor of at least 80 times lower intensity than in the $^{12}$CO line wings, the $^{12}$CO wing emission is most likely optically-thin. Higher resolution (20″) followup observations of the central region in the CO $J=3\rightarrow2$ line shown in Fig. 4 confirm the presence of a compact outflow. In this transition, which generally traces warmer gas than the $J=2\rightarrow1$ transition, the emission peaks are separated by 8±1″ along a position angle +19°.

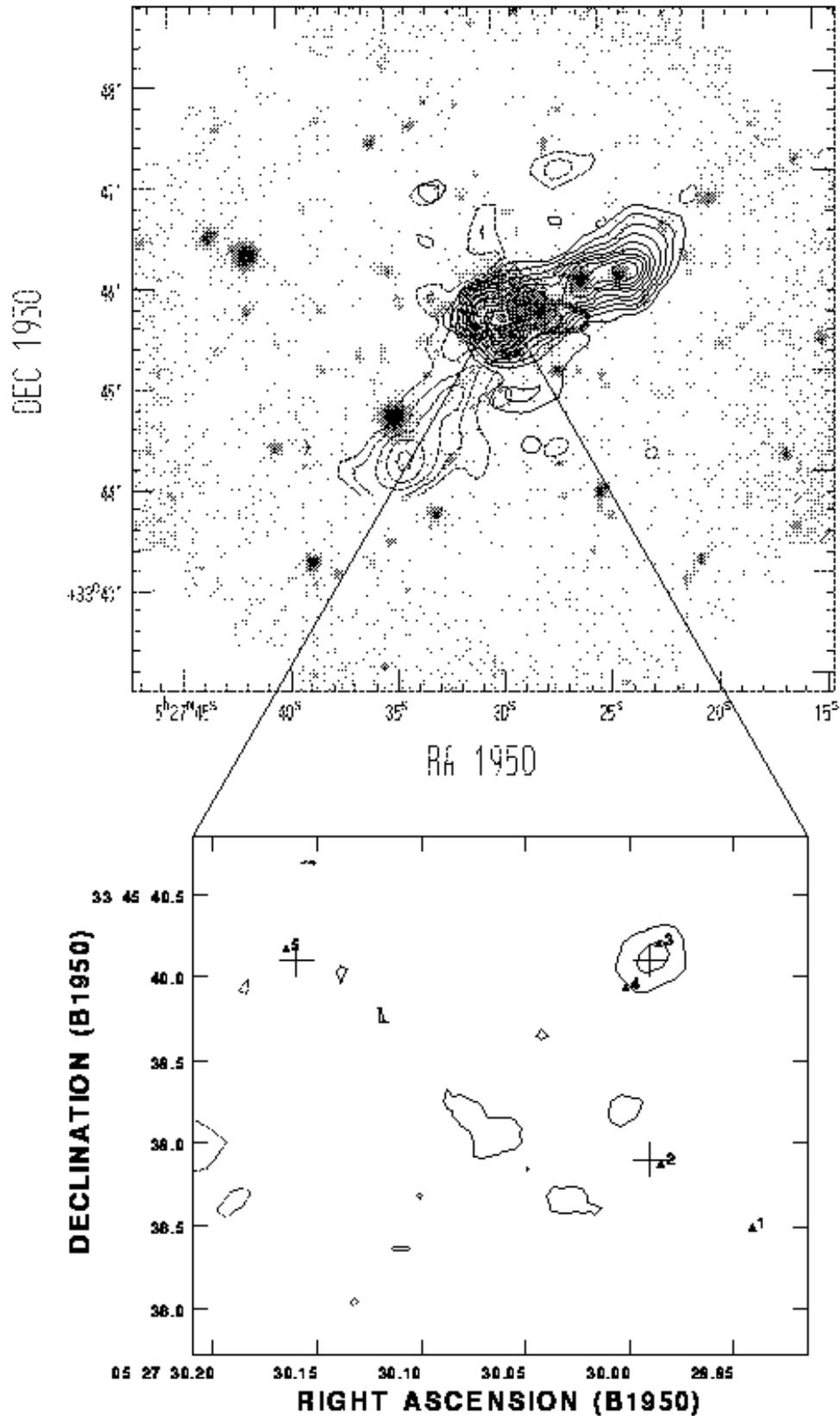

**Fig. 2.** Upper panel: $7' \times 7'$ K–band mosaic. The greyscale is logarithmic. The ellipse represent the error box of IRAS 05274+3345. Superimposed on the greyscale is the CO $J=2\to1$ outflow map with solid and dashed contours indicating blueshifted (-18 to -8 km s$^{-1}$) and redshifted (2 to 12 km s$^{-1}$) emission (2 to 10 by 1 K km s$^{-1}$). Lower panel: the 8.4 GHz contour map with the maser positions marked by triangles (crosses mark Torrelles et al. 1992 positions.)

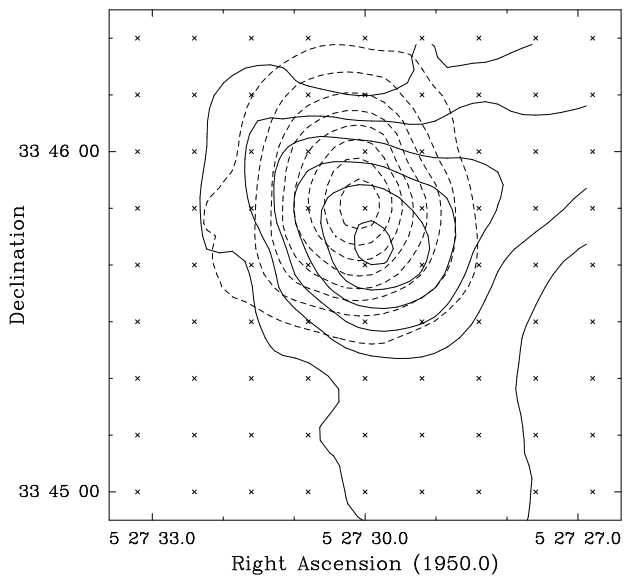

**Fig. 4.** A contour map of the CO $J=3\rightarrow2$ emission with thte blueshifted emission (-18 to -8 km s$^{-1}$) shown in solid contours and the redshifted emission (2 to 12 km s$^{-1}$) shown in dashed contours (levels 8 to 64 by 8 K km s$^{-1}$). The observed positions are marked by crosses.

### 3.2. Dense molecular core

With an excitation temperature of 66 K and a critical density of $3.2 \times 10^7$ cm$^{-3}$ (Green & Chapman 1978), the CS $J=7\rightarrow6$ transition traces warm, dense gas. In AFGL 5142, the peak of the CS emission occurs at the center of the compact outflow. Based on our map, the peak is unresolved, indicating that the high density molecular gas is concentrated in a core smaller than the beamsize (20″∼ 0.17 pc). In addition, as shown in Fig. 1, the broad wings visible in the CS spectrum provide evidence that the dense gas in the inner region of the outflow is probably following the flow traced at a larger scale by the compact CO outflow; hence the compact outflow is active also on a very small scale.

Due to the broadening of the line, the CS linewidth cannot be used to obtain an accurate value of the virial mass of the core. Instead, we use the linewidths of $^{13}$CO $J=2\rightarrow1$ and $6\rightarrow5$. Based on the full-widths half-maximum of the $^{13}$CO $J=2\rightarrow1$ and $6\rightarrow5$ lines of 3.48±0.07 km s$^{-1}$ and 4.05±0.08 km s$^{-1}$, we estimate a virial mass within (the observed beamsize of) 16.5″ radius (0.14 pc) of 400±20 $M_\odot$ and within 5.5″ radius (0.048 pc) of 180±20 $M_\odot$. Assuming a spherically symmetric density profile, these two data points are consistent with a profile of $n(r) \sim r^{-2.3}$.

### 3.3. Water masers and radio continuum

The contour map in the lower panel of Fig. 2 shows the 8.4 GHz continuum from the inner core of AFGL 5142. The observed H$_2$O masers are marked by filled triangles and numbered. We identify a weak continuum barely-resolved source with peak flux density of 0.50±0.09 mJy and integrated flux of 0.83±0.15 mJy at $\alpha = 05^h27^m29\overset{s}{.}989$, $\delta = +33° 45'40\overset{''}{.}110$ nearly aligned with maser spots C3 and C4. The signal to noise ratio of 6 corresponds to an absolute position uncertainty of ±0.03″ and confirms the VLA C-array detection at the same frequency of a 0.69 mJy source by Torrelles et al. (1992) and the 5 GHz detection of McCutcheon et al. (1991) and Carpenter et al. (1990).

The importance of our VLA A-array observations is that they show that all the flux density (within the relative uncertainties) is contained in a barely-resolved source at 0.3″ resolution. We thus exclude the presence of a cometary H II region (evidenced in Torrelles et al. (1992) only by the lower contours). The spectral index between 5 and 8.4 GHz is $\alpha \sim 1$, consistent, within the uncertainty, with that of an ionized wind (Panagia & Felli 1975). This source was undetected in our observations at 22 GHz which have an rms level of 1.9 mJy/beam. Unfortunately, this limit is too high to confirm or disprove the wind hypothesis, since for a spectral index $\alpha = 0.6$ the expected wind flux density at 22 GHz would be $\sim 1.2$ mJy, i.e. well below the noise. However, an optically thick H II region (spectral index $\alpha = 2$) would give an expected flux density of $\sim 5$ mJy, thus this possibility is less likely.

Given the spectral index and the very small angular size, we favor the interpretation of the continuum emission in terms of an ionized stellar wind rather than in terms of a standard (i.e constant density spherical) UC H II region. In fact, the high density (and high internal pressure) implied in the UC H II region hypothesis would predict a very short expansion life-time (unless other confining agents such as blisters, or bow-shocks are present), thus reducing the chance probability of observing such a rapidly evolving stage. Instead, ionized stellar winds have inherent small diameters (Panagia & Felli 1975) and do not suffer the above mentioned shortcoming. In the hypothesis of a radiatively ionized stellar wind the free-free emission is no longer optically thin and the required stellar ionizing flux can be much larger than that derived in the optically thin approximation. Consequently, the spectral type of the exciting star can be earlier than B2.

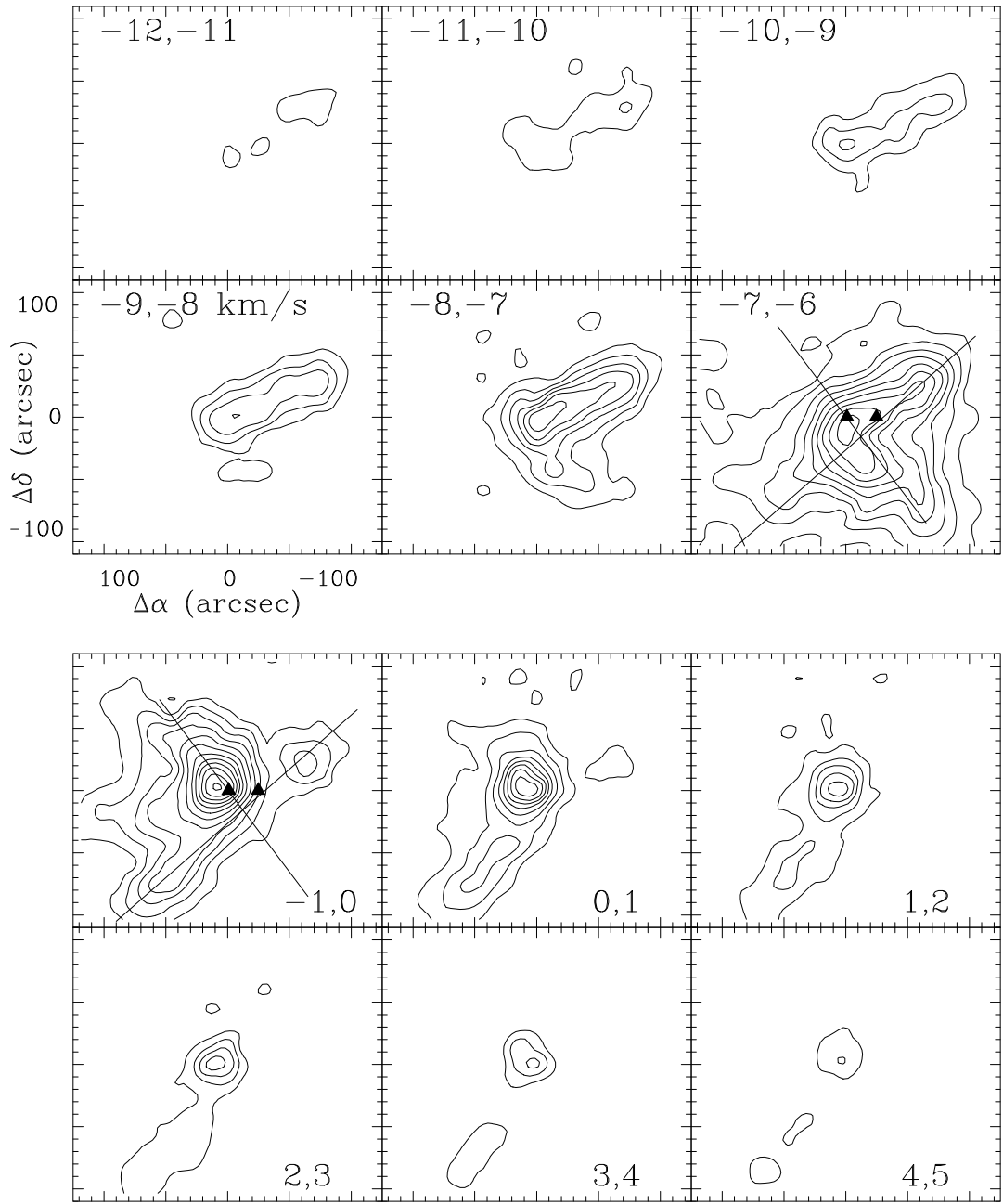

**Fig. 3.** Channel maps of the CO 2-1 emission wings (one K km s$^{-1}$ each) with contours beginning at 0.75 with increments of 0.75 K km s$^{-1}$. The offsets are from: $\alpha = 05^h27^m30\overset{s}{.}0$  $\delta = 33° 45' 40\overset{''}{.}0$. The position of the NIR sources IRS1 and IRS2 are plotted for reference. The suggested axes of the two outflows are also marked.

The position, flux and isotropic luminosity of the maser components are listed in Table 1; the corresponding spectra are presented in Fig. 5, together with a single dish spectrum taken with the Medicina radiotelescope (Tofani et al. 1995) two months after the VLA run.

The bulk of the $H_2O$ emission occurs at velocities near the quiescent molecular cloud velocity ($\sim$ -4.1 km s$^{-1}$) and within the inner boundaries (from -8 to 2 km s$^{-1}$) that define the two outflow lobes. No emission is found outside this interval in the Medicina spectra which cover a much larger bandwidth than that shown in Fig. 5. This confirms that the maser formation occurs in the inner region of the outflow, where the acceleration of the molecular gas has just begun.

Maser components C3 and C4 are located very close to the continuum source and on opposite sides of it, at a distance from the continuum source of 130±50 and 350±50 AU, respectively. These two components have broad spectra, similar to the masers embedded in the UCH II region W75N (Hunter et al. 1994) and are markedly offset in velocity. C3 extends on the red side of -2 km s$^{-1}$, while C4 is on the blue side of 0 km s$^{-1}$. This indicates a drastic change of the velocity pattern in the immediate surroundings and on opposite sides of the YSO.

Component C1 is the weakest maser component and probably was undetected in the Torrelles et al. (1992) study, possibly because it has a highly variable flux. Components C2 and C5, the next two most distant spots from the YSO ($\sim$2300 and 3900 AU, respectively), have velocities very close to the quiescent molecular cloud and most probably represent masers moving in the plane of the sky. They lie near the respective centroids of two Torrelles et al. (1992) spots (marked with a cross in Fig. 2), but show an offset with respect to their positions. The position offsets between the two epochs ($\sim$ 1.8 yr baseline) can be interpreted as possible evidence for proper motion. Both components seem to move outward from the center position: C2 has moved southeast at 38±50 mas yr$^{-1}$ while C5 has moved northwest at 49±50 mas yr$^{-1}$. Clearly the errors (based on the absolute positional uncertainties given by Torrelles et al. 1992) are as large as the putative proper motion and we take this only as an indication that should be confirmed by further VLA A-array observations. However, if confirmed these proper motions imply projected velocities of $\sim$370 ±300 km s$^{-1}$ along a +60° axis–closer in alignment with the compact outflow than the extended outflow. Although velocities of this order of magnitude ($\geq$ 100 km s$^{-1}$) have been previously reported in Orion-KL (Genzel et al. 1981) and in W49N (Gwinn et al. 1992), they are inconsistent with the observed velocity of the line emission, unless both maser components happen to be moving very nearly in the plane of the sky. In both cases components with high proper motions are found only at large distances from the YSO (for Gwinn et al. (1992) the acceleration takes place in a 300 AU region at $\sim 10^4$ AU from the YSO), in agreement with what we find in AFGL5142.

The $H_2O$ emission is highly variable. For instance the Medicina spectrum does not show any feature at $-10$ km s$^{-1}$, where C4 presents the brightest peak. Consequently, this velocity component must have disappeared on time scales of only a couple months. On a longer time scale (years) the information available from the literature (Verdes-Montenegro et al. 1989, Felli et al. 1992[4], Brand et al. 1994, Delin & Turner 1993) indicates that: 1) the emission at positive velocities had a peak in the late 1980's and then continuously decreased, 2) in the 1990's the more stable component was that at $\sim$ -5 km s$^{-1}$ , and 3) velocity components between -10 and 0 km s$^{-1}$ appear and disappear with time scales of the order of the sampling time (approximately one year). This large variability on short time–scales indicates energetic activity within a few thousand AU around the YSO. An undesirable consequence is that any direct comparison between maser features and more permanent structures like the $H_2$ jets and the CO outflow is highly unreliable.

### 3.4. The young stellar cluster

In Fig. 2 the 7'×7' K–band mosaic is presented. Over 250 sources have been detected in the field, but for only 100 of them is it possible to estimate all three J, H and K magnitudes. Compared to other regions of the same type (see for example Palla et al. 1995, Testi et al. 1994), this field seems to have a low spatial density of sources. However, it is evident that there exists a central cluster embedded in a faint nebulosity.

To check quantitatively the clustering around the position of the radio continuum source we calculate the source densities at various radii.

In Fig. 6, the K–Band source density is plotted as a function of the projected distance from the radio continuum source. The density is calculated at radii ranging from 12" to 180" with increments of 12". At each radius $r_i$ the density is calculated as the number of sources detected between $r_{i-1} = r_i - 12$ " and $r_i$, divided by the area $A = \pi(r_i^2 - r_{i-1}^2)$. The density of sources shows an obvious peak at the central position with a width of $0.3 - 0.4$ pc; for $r > 0.6$ pc the density drops to an almost constant value ten times lower than that at the center, representing the background star density. Similar curves calculated with the central position shifted show reduced maxima and are broader, implying that the center of the cluster is within $\sim 10''$ from the position of the radio continuum source. Therefore, IRS1 lies closer than IRS2 to the center of the cluster.

---

[4] the velocity scale of the spectrum published in Felli et al. 1992 and the velocities given in their table 2 for this component should be diminished by 20 km s$^{-1}$due to an offset of 1.4 MHz discovered afterwards in the LO

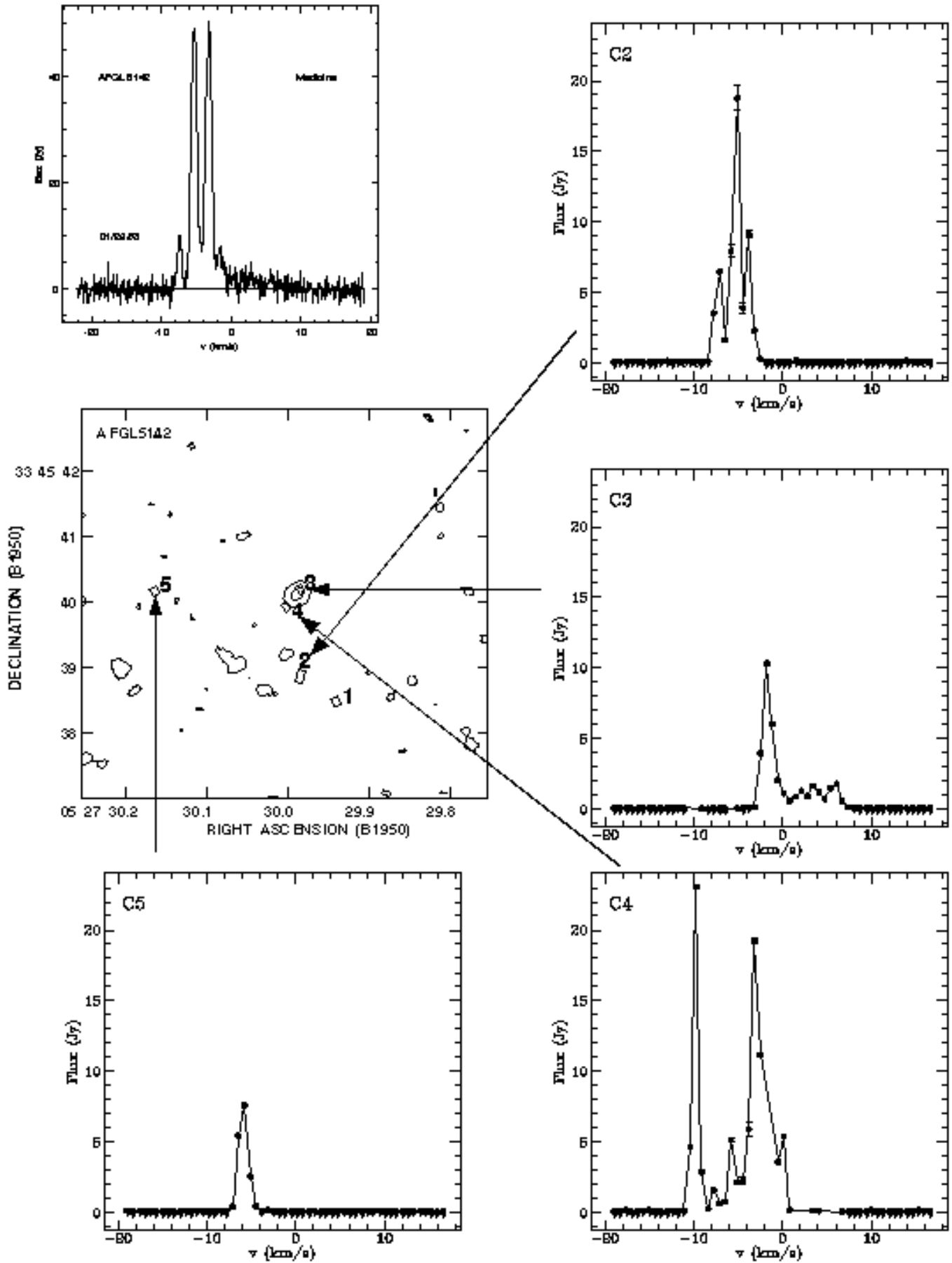

**Fig. 5.** The VLA spectra of the brightest maser spots are shown. For comparison the Medicina single dish spectrum taken after the VLA run is also shown.

**Table 1.** AFGL 5142 $H_2O$ Maser Components.

| Comp. | Coordinates (1950) R.A. | Decl. | Peak Flux (Jy) | Integ. Flux (Jy km s$^{-1}$) | LSR Vel. (km s$^{-1}$) $v_{peak}$ | $v_{min}$ | $v_{max}$ | Luminosity ($L_\odot$) | Dist. from YSO (AU) |
|---|---|---|---|---|---|---|---|---|---|
| 1 | $05^h27^m29^s.941$ | $33°45'38''.487$ | $0.62 \pm 0.09$ | $0.64 \pm 0.07$ | -8.3 | -9 | -8.3 | $2.9 \times 10^{-8}$ | 3100 |
| 2 | 05 27 29.985 | 33 45 38.871 | $18.8 \pm 1.29$ | $35.7 \pm 1.09$ | -5 | -7.7 | -2.4 | $1.6 \times 10^{-6}$ | 2300 |
| 3 | 05 27 29.986 | 33 45 40.196 | $10.3 \pm 0.53$ | $23.4 \pm 0.49$ | -1.7 | -9 | 7.4 | $1.0 \times 10^{-6}$ | 130 |
| 4 | 05 27 30.002 | 33 45 39.928 | $23.1 \pm 1.16$ | $59.0 \pm 1.21$ | -9.6 | -10.3 | 6.7 | $2.6 \times 10^{-6}$ | 350 |
| 5 | 05 27 30.165 | 33 45 40.164 | $7.63 \pm 0.39$ | $11.1 \pm 0.34$ | -5.7 | -7 | -3.1 | $5.0 \times 10^{-7}$ | 3900 |

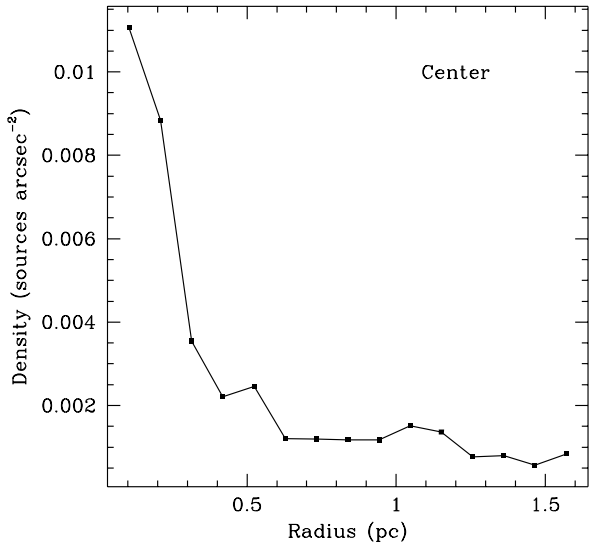

**Fig. 6.** K–Band source density as a function of the distance from the radio continuum source. The horizontal scale is the projected linear radius on the plane of the sky assuming a distance of 1.8 kpc (1 pc= 115″).

The result that young stars with infrared excess tend to be confined in sub–parsec scale clusters has been obtained also for other $H_2O$–maser/CO–outflow regions studied, such as the BD+40° 412 (Palla et al. 1995) and NGC 3576 (Persi et al. 1994). The cluster sizes are 0.35 pc in AFGL 5142, 0.19 in the BD+40° 412 region and 0.58 pc in NGC 3576.

Contour plots of the cluster in the three J, H and K broad bands are shown in Fig. 7, the region covered by the cluster is $1.5' \times 1.5'$ wide. In Table 2 absolute positions and magnitudes of the sources detected in this field are given. When the quoted magnitude is a lower limit it is indicated with a > symbol. Unless otherwise noted, the quoted photometric errors are 0.1 magnitudes. The source labelled IRS1 is coincident, within our position uncertainties, with the radio continuum source detected at 8.4 GHz. The source labelled IRS2 is the brightest source and nearly coincident with the center of the error ellipse of the far infrared source IRAS 05274 + 3345.

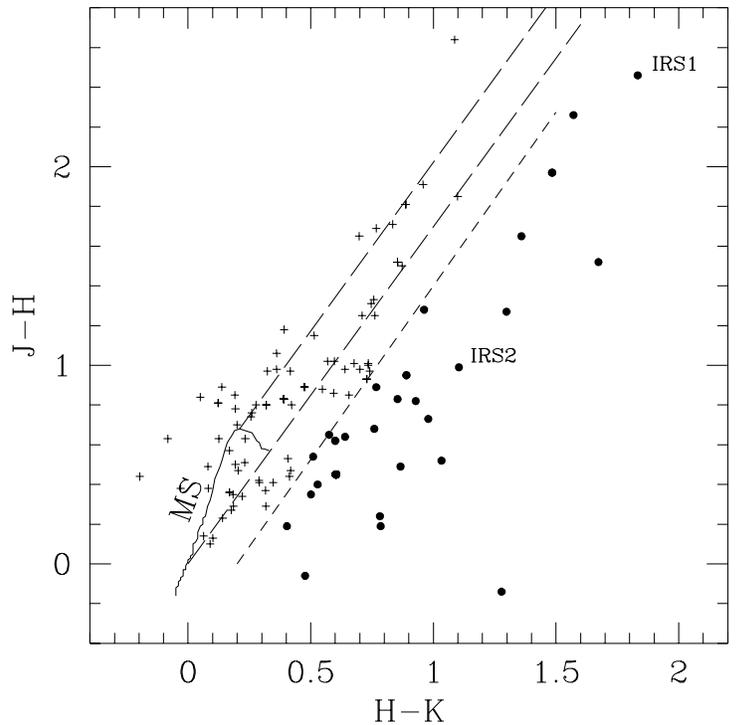

**Fig. 8.** (H−K, J−H) color–color plot for the sources with good measurements in the three broad bands. Filled circles represent sources with infrared excess, as defined in the text, crosses sources without. Solid line labelled MS marks the position of unreddened main sequence stars, the two long dashed lines the position of the reddening belt for main sequence stars. The short dashed line is the line used to discriminate stars with and without infrared excess.

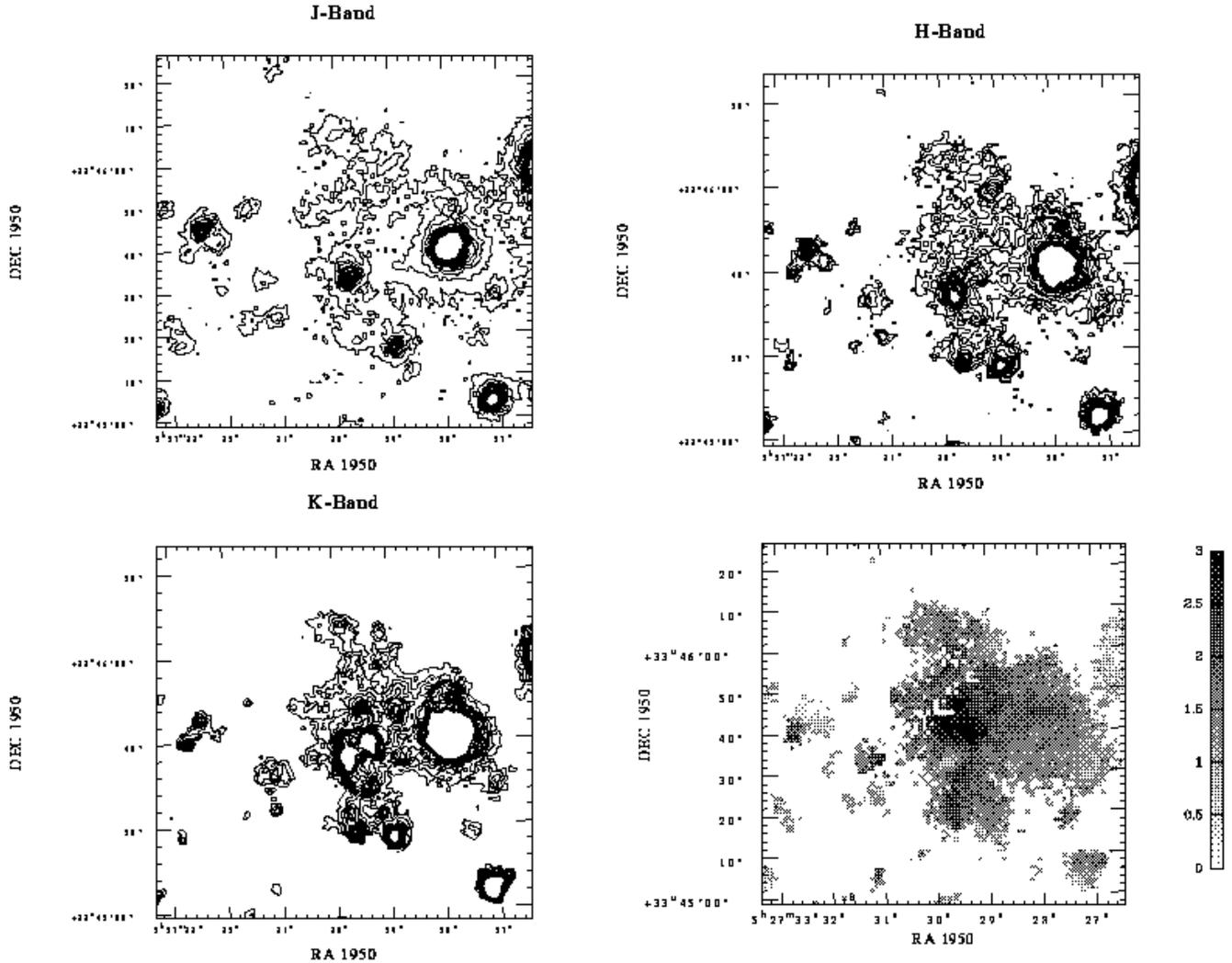

**Fig. 7.** From left to right: contour plots of the central cluster in J, H and K and greyscale image of the J − K color index map. The lowest contour correspond to 0.01, 0.02 and 0.03 milli–Jansky per square arcsec in the three bands respectively. The other contours are spaced by 0.01 milli-Jansky per square arcsecond.

In Fig. 8 the (H − K, J − H) color–color diagram for all 100 sources with good measurements is presented. As discussed in Lada & Adams (1992), this diagram provides a powerful tool to identify different stages of evolution of stellar objects, even though the J, H and K near infrared colors alone may not allow one to draw definitive conclusions on low mass objects (Aspin & Barsony 1994).

We define a line ([J − H] = 1.75 × [H − K] − 0.35) parallel to the reddening law on the color–color plane in order to separate qualitatively the sources with infrared excess from the unreddened main–sequence and reddened main–sequence stars. This line takes into account 1-sigma uncertainties in the observed magnitudes. In principle all the stars that lie on the right side of the reddening belt in Fig. 8 should have an infrared excess, but the scatter of the measurements around the main sequence may produce spurious infrared excess for the sources near the reddening belt. To avoid this effect all the sources on the right of the defined line should have a bona–fide infrared excess. Hence the number of stars with infrared excess that we measure is a lower limit to the real value in the field. Even with this restriction, a large fraction of the sources show a near infrared excess (∼ 29%).

The clustering of sources around the central position is even more evident when looking only at sources which show infrared excess. In fact, in the central region the fraction of sources with infrared excess is higher than the mean value for the entire region. Out of a total of 25 sources listed in Table 2 with good measurements in all the three bands, 11 (44%) show infrared excess, with the criterion described before. IRS1 is the source with the highest color index, while IRS2, even if it exhibits an infrared excess, is not the "reddest" source of the cluster (see also Fig. 8).

**Table 2.** Sources detected in the cluster region.

| # | $\alpha$ (1950) | $\delta$ (1950) | $m_J$ | $e_J$ | $m_H$ | $e_H$ | $m_K$ | $e_K$ | name |
|---|---|---|---|---|---|---|---|---|---|
| 1 | 5 : 27 : 27.13 | 33 : 45 : 28.7 | 15.7 | | 15.0 | | 14.3 | | |
| 2 | 5 : 27 : 27.23 | 33 : 45 : 03.8 | 13.9 | | 13.0 | | 12.2 | | |
| 3 | 5 : 27 : 27.55 | 33 : 45 : 17.8 | 17.4 | (.26) | > 16.5 | | 14.8 | (.13) | |
| 4 | 5 : 27 : 27.89 | 33 : 45 : 48.9 | 15.3 | | 13.9 | | 13.2 | | |
| 5 | 5 : 27 : 28.01 | 33 : 45 : 40.1 | 12.6 | | 11.6 | | 10.5 | | IRS2 |
| 6 | 5 : 27 : 28.94 | 33 : 45 : 46.5 | 16.7 | (.17) | 15.1 | (.15) | 13.7 | | |
| 7 | 5 : 27 : 29.01 | 33 : 45 : 16.9 | 15.1 | | 13.8 | | 13.1 | | |
| 8 | 5 : 27 : 29.10 | 33 : 45 : 59.0 | 16.2 | (.11) | 15.4 | (.19) | 14.7 | (.28) | |
| 9 | 5 : 27 : 29.15 | 33 : 45 : 22.3 | 16.3 | | 14.8 | | 13.9 | | |
| 10 | 5 : 27 : 29.17 | 33 : 46 : 05.8 | 17.1 | (.2) | 15.5 | (.17) | 14.8 | (.20) | |
| 11 | 5 : 27 : 29.41 | 33 : 45 : 38.8 | 16.7 | (.13) | 14.4 | | 12.9 | | |
| 12 | 5 : 27 : 29.46 | 33 : 45 : 28.3 | 16.8 | (.19) | 14.8 | (.11) | 13.4 | | |
| 13 | 5 : 27 : 29.53 | 33 : 45 : 47.1 | > 17.5 | | 15.9 | (.26) | 13.8 | | |
| 14 | 5 : 27 : 29.71 | 33 : 45 : 17.8 | 16.3 | | 14.5 | | 13.6 | | |
| 15 | 5 : 27 : 29.75 | 33 : 45 : 56.2 | > 17.5 | | > 16.5 | | 14.5 | (.16) | |
| 16 | 5 : 27 : 29.80 | 33 : 45 : 22.9 | 16.8 | (.14) | 14.9 | | 13.8 | | |
| 17 | 5 : 27 : 29.83 | 33 : 45 : 32.9 | 14.7 | | 13.9 | | 13.0 | | |
| 18 | 5 : 27 : 29.91 | 33 : 45 : 36.2 | 15.4 | | 13.7 | | 12.9 | | |
| 19 | 5 : 27 : 29.92 | 33 : 45 : 40.2 | 17.5 | (.3) | 15.1 | (.14) | 13.2 | | IRS1 |
| 20 | 5 : 27 : 29.97 | 33 : 46 : 07.1 | 16.1 | | 14.7 | | 14.0 | | |
| 21 | 5 : 27 : 31.09 | 33 : 45 : 32.6 | > 17.5 | | 15.8 | (.15) | 14.1 | (.12) | |
| 22 | 5 : 27 : 31.17 | 33 : 45 : 24.1 | 16.4 | | 15.4 | (.12) | 14.8 | | |
| 23 | 5 : 27 : 31.23 | 33 : 45 : 04.2 | 17.7 | (.2) | 16.5 | (.23) | 15.2 | | |
| 24 | 5 : 27 : 31.45 | 33 : 45 : 32.7 | 16.8 | (.12) | 15.3 | | 14.5 | | |
| 25 | 5 : 27 : 31.67 | 33 : 45 : 50.1 | 16.5 | | 16.1 | (.21) | > 16 | | |
| 26 | 5 : 27 : 31.70 | 33 : 45 : 22.2 | 16.9 | (.11) | 15.9 | (.17) | 15.3 | (.14) | |
| 27 | 5 : 27 : 32.20 | 33 : 45 : 41.9 | 15.9 | | 15.1 | | 14.8 | | |
| 28 | 5 : 27 : 32.52 | 33 : 45 : 45.3 | 14.8 | | 14.4 | | 14.1 | | |
| 29 | 5 : 27 : 32.81 | 33 : 45 : 39.7 | 17.1 | (.13) | 15.2 | | 14.2 | | |
| 30 | 5 : 27 : 32.90 | 33 : 45 : 18.0 | 16.4 | | 15.5 | (.11) | 15.2 | | |

Note: The limiting magnitudes are 17.5, 16.5 and 16.0 in J, H and K.

The J−K color index map of the cluster region (Fig. 7), shows that the highest values of the color index are associated with the IRS1 region. Since both IRS2 and the diffuse region surrounding it have similar, lower values of the color index, the diffuse emission is likely to be a reflection nebula. Somewhat surprisingly the peak of the IRAS source does not coincide with the source exhibiting the highest infrared excess (IRS1), the $H_2O$ masers and the continuum radio source, but with the brightest source (IRS2) which also (most probably) illuminates the diffuse emission. This positional distinction between IRAS sources and NIR sources with high color excess has been found in other regions of the same type (Testi et al. 1994). We conclude that the IRAS source traces the region of the extended nebulosity, usually associated with the most luminous source.

### 3.5. Narrow band images

In Fig. 9 contour plots of the Brackett $\gamma$ images of the central cluster are shown. On the left is the image before convolution and subtraction; on the right is the convolved image with the synthetic continuum subtracted.

Obviously the subtraction procedure has failed to completely remove the IRS2 source. Unfortunately ARNICA does not have continuum narrow band filters adjacent to the lines which would improve the subtraction. In spite of this fact the subtraction procedure successfully removes most of the stars that are likely to have no line emission or absorption detectable; thus the applied algorithm is reliable for our purposes.

The Brackett $\gamma$ image does not show any line emission feature, even at the position of the radio continuum source (coincident with IRS1), down to $\sim 1.6 \times 10^{-15}$ erg s$^{-1}$ cm$^{-2}$ arcsec$^{-2}$. For the emission associated with IRS1, taking into account the continuum subtraction noise, we can estimate an upper limit to the Brackett $\gamma$ flux of $\sim 10^{-14}$ erg s$^{-1}$ cm$^{-2}$. This value can be used to estimate a lower limit to the extinction to IRS1 in the case that the radio continuum emission comes from a stellar wind. In fact, in this hypothesis (Simon et al. 1983): $F(Br\gamma) \sim 1.6 \times 10^{-12} S_{8.5\,GHz}(mJy)$ erg s$^{-1}$ cm$^{-2}$. From the measured radio continuum flux we would obtain $F(Br\gamma) \sim 1.3 \times 10^{-12}$ erg s$^{-1}$ cm$^{-2}$, hence, for the visual extinction toward the source we find $A_V \geq 50$.

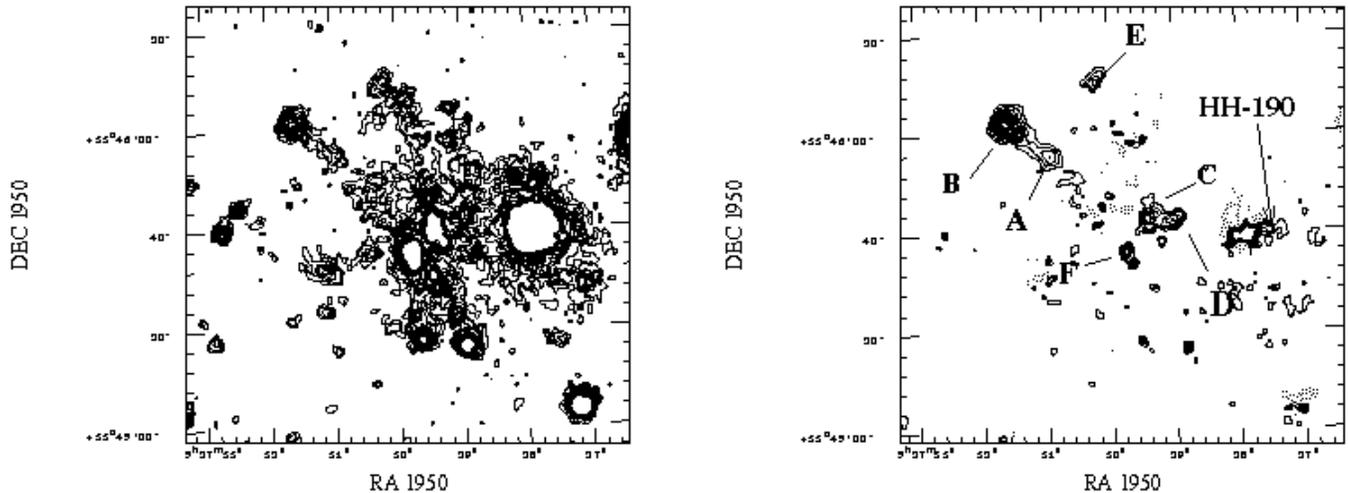

**Fig. 9.** Contour plots of the $H_2$ image. On the left: before convolution and continuum subtraction. On the right: after convolution and continuum subtraction. Solid contours: positive values; dashed contours: negative values. The first contour (positive and negative) corresponds to three times the noise ($\sim 1.6 \times 10^{-16}$ erg cm$^{-2}$ s$^{-1}$ ″$^{-2}$), all the other contours are spaced by $1.6 \times 10^{-16}$ erg cm$^{-2}$ s$^{-1}$ ″$^{-2}$. The detected knots are labelled for reference with Table 3. The position of the HH object discovered by Eiroa et al. (1994) is also marked.

On the other hand, the $H_2$ image shows a rather complex line emission structure. Even before the continuum subtraction, when compared to the K–Band image of Fig. 7, a jet–like structure is evident to the north–east of IRS1. After continuum subtraction a number of other knots appear more clearly. In Fig. 9 the knots are labelled from A to F and in Table 3 their position and fluxes are reported. Since the jet–like structure does not show any continuum emission, the fluxes of components A and B are more accurate (8–10%). For all the other clumps the flux estimates are less accurate due to continuum subtraction noise ($\sim 20\%$). The F component has a measured flux that is of the order of three times the noise induced by the continuum subtraction and hence is only marginally detected.

**Table 3.** Integrated fluxes of the $H_2$ knots.

| Knot | Coordinates (1950) R.A. | Decl. | Dist.[a] (pc) | Int. Flux (erg cm$^{-2}$ s$^{-1}$) |
|---|---|---|---|---|
| A | 5 : 27 : 30.91 | 33 : 45 : 55.6 | 0.172 | $2.6 \times 10^{-14}$ |
| B | 5 : 27 : 31.60 | 33 : 46 : 02.6 | 0.268 | $5.4 \times 10^{-14}$ |
| C | 5 : 27 : 29.41 | 33 : 45 : 42.4 | 0.059 | $3.5 \times 10^{-14}$ |
| D | 5 : 27 : 29.02 | 33 : 45 : 42.3 | 0.100 | $4.0 \times 10^{-14}$ |
| E | 5 : 27 : 30.17 | 33 : 46 : 10.6 | 0.267 | $1.6 \times 10^{-14}$ |
| F | 5 : 27 : 29.74 | 33 : 45 : 35.7 | 0.044 | $3.1 \times 10^{-14}$ |

[a] Distances are measured from IRS1

The jet–like feature A–B is pointing outward from IRS1 (see Fig. 10) at a position angle $\sim 46°$, and the possible counter–jet C–D at $\sim -72°$. The eastern component of the jet is aligned to within few degrees of the position angle of the compact outflow discussed in section 4.1, see also Fig. 3. This almost perfect alignment between the $H_2$–jet and the compact molecular outflow suggests that they are physically related.

## 4. Discussion

### 4.1. The nature of two outflows

The CO $J=2\to1$ channel maps (Fig. 3, clearly indicate a large scale bipolar outflow but with additional structure in the perpendicular direction at the origin. There are three possible interpretations for this smaller scale structure: 1) a rotating circumstellar disk collimating the large scale outflow; 2) part of a single poorly-collimated conical outflow in which only the limb-brightened edges are seen; and 3) a more compact outflow, unrelated to the large scale one. For several reasons, we favor the explanation of a more compact outflow near the origin of the extended flow. First, the structure is too large to be considered a circumstellar disk since the virial mass required to bind it ($M \sim R\bar{v}^2/G \sim 2000 M_\odot$) is nearly three orders of magnitude larger than the measured mass and a factor of 5 larger than the virial mass of the cloud core estimated from the $^{13}$CO $J=2\to1$ linewidth. Second, the observed geometry of the lobes–the large difference in their lengths and their nearly perpendicular alignment–requires a large opening angle ($\sim 90°$) coupled with asymmetric expansion if it is to be explained by a single edge-brightened outflow. Third, our discovery of a near infrared H$_2$ jet in the same direction of the offset and on a scale size of $\sim$ 20″(0.17 pc) supports the existence of a more compact outflow associated with IRS1. Also, the presence of multiple outflows with different size scales in the same starforming region has been reported or speculated in several cases (e.g. DR21 Garden et al. 1991, and Mon R2 Wolf et al. 1990). For these reasons, we believe the presence of two outflows to be the simpler and more likely interpretation of the CO emission.

Proceeding with the assumption of two distinct outflows, the emission in the CO channel maps has been spatially partitioned into two components: an extended outflow in the NW/SE direction, and a compact outflow in the NE/SW direction. Emission from the 73 sampled positions closest to the NE/SW axis of the map was assigned to the compact outflow while emission from the rest of the 182 sampled positions was assigned to the extended outflow. The column density, mass and energetics of the outflowing gas were computed using the beam-filling factor approach detailed by Garden et al. (1991) assuming CO to be optically-thin based on the $^{13}$CO data. The derived parameters of the outflows are presented in Table 4. The characteristic velocity, $\bar{v}$, is the intensity-weighted radial velocity measured from line center (-3.5 km s$^{-1}$) for the redshifted (-1 to 12 km s$^{-1}$) and blueshifted (-6 to -18 km s$^{-1}$) emission. An estimate of the dynamical timescale is derived from the physical length divided by the characteristic velocity. The mass ejection rate is then computed by dividing the total mass by the dynamical timescale. In the channel maps nearest the LSR velocity, the extended outflow shows emission in both lobes, indicating a large inclination angle. We have estimated the inclination of the two outflows using the method described by Cabrit & Bertout (1990) and computed the geometrically-corrected values of the physical parameters. For the extended outflow, spectra at the outer emission peaks, offsets (-75″,+30″) and (+60″,-90″) relative to the map center, indicate that the CO line emission is nonzero over the velocity ranges: -21.0 to +0.5 km s$^{-1}$, and -7.0 to +14.0 km s$^{-1}$, respectively. These values yield an average ratio ($R$) of 4.69 relative to the LSR velocity (-3.5 km s$^{-1}$). Based on the half-power width at the end of the outflow (31″) and the length (105″), the maximum opening half-angle of the outflow is estimated to be 8.4°. In the formula of Cabrit & Bertout (1990), these numbers imply an inclination of 84.5° (nearly in the plane of the sky). A similar analysis was carried out for the compact outflow, but only the spectrum at offset (+15″,+15″) was considered due to contamination of the blueshifted wing by the extended outflow. Here the emission extends from -12.0 to +14.0 km s$^{-1}$ for a ratio of 2.06, and based on the width (43″) and length (40″), the maximum opening half-angle of the outflow is 28.3°. These parameters imply an inclination of 79.5°. The correction factors applied in Table 4 are quite large because both outflows present high inclinations; thus the corrected parameters should be considered only as approximate values. Based on the inclination-corrected parameters, it is remarkable that in spite of the difference in the observed length of the two outflows (the large scale outflow is nearly a factor of 3 longer than the compact one), the timescales and, in turn, the mass ejection rates, are nearly identical. This agreement suggests that the basic driving mechanism is the same for both outflows. The gas in the extended outflow is simply being ejected at a higher velocity, perhaps because its origin lies further from the center of the dense core of the molecular cloud. In support of this interpretation, the axis of the two outflows, as shown in Fig. 3, seem to indicate that there are two centers displaced by several arcseconds: the compact outflow is clearly centered on IRS1, while IRS2 lies almost exactly on the axis of the extended outflow.

The region around IRS1/IRS2 has been recently observed by Eiroa et al. (1994) in the I broad band (0.9 $\mu$m) and in the H$\alpha$ and S[II] optical narrow bands, revealing the presence of several red stars and an Herbig–Haro object (HH 190). The broad band CCD image (Fig. 2 in their paper) shows a cometary shaped reflection nebulosity to the west of their "star–1", which is coincident with our IRS2 source. No diffuse emission is evident on the east side of the source and no emission is detected at the position of IRS1 and of the other peaks in the (J−K) color index map of Fig. 7. In the narrow band images they detect the presence of HH 190 located 7″ to the west of IRS2, but they do not detect any emission near IRS1 where we found the H$_2$ jet–like structure. This is probably due to the fact that the region around IRS1 is more heavily extincted. The position of HH 190 is marked in Fig. 9. In the H$_2$ image there is a small excess of emission at the position of the HH object, but the subtraction noise induced by the presence of the bright IRS2 source a few arcseconds away makes any estimate of the flux extremely uncertain.

The comparison between our infrared images and the optical CCD frames confirms the presence of a cluster of young embedded sources. The fact that the region around IRS2 is less extincted than that around IRS1 suggests that IRS1 lies deeper inside the molecular clump, while IRS2 may be at the edge of it, consistent with the interpretation from the CO outflow maps. The extended CO outflow is aligned with IRS2, suggesting that this young star could be the powering source of the flow. The compact outflow is, however, centered at the position of IRS1, and it is aligned with the H$_2$ jet axis and the marginally detected H$_2$O maser spots proper motion. Thus we believe the extended outflow has been driven by the processes of the formation of IRS2, while the compact outflow is being driven by the IRS1 source.

Regarding the FIR emission detected by IRAS, we expect that IRS2 dominates the mid infrared emission from the region, while the most embedded sources are probably responsible for the bulk of the emission in the 60 and

Table 4. Derived Parameters of the Bipolar Outflows in AFGL 5142.

| Wing | Component | $\bar{v}$ ( km s$^{-1}$) | Length ($''$) | Length (pc) | Timescale ($10^5$ yr) | Mass ($M_\odot$) | Momentum ($M_\odot$ km s$^{-1}$) | Luminosity ($L_\odot$) | Mass Loss ($10^{-5} M_\odot$ yr$^{-1}$) |
|---|---|---|---|---|---|---|---|---|---|
| Red | Extended | 5.7 | 120 | 1.05 | 1.6 | 1.6 | 9.3 | 0.036 | 1.0 |
| Blue | Extended | 4.6 | 95 | 0.79 | 2.0 | 1.5 | 6.9 | 0.017 | 0.7 |
| Total | Extended | 5.2 | 215 | 1.84 | 1.8 | 3.1 | 16.2 | 0.053 | 1.7 |
| Corrected | $i = 84.5°$ | 54 | 217 | 1.86 | 0.17 | 3.1 | 170 | 5.8 | 18 |
| Red | Compact | 5.2 | 40 | 0.35 | 0.66 | 1.2 | 6.0 | 0.053 | 1.8 |
| Blue | Compact | 4.6 | 40 | 0.35 | 0.75 | 1.2 | 5.5 | 0.036 | 1.6 |
| Total | Compact | 4.9 | 80 | 0.70 | 0.70 | 2.4 | 11.5 | 0.089 | 3.4 |
| Corrected | $i = 79.5°$ | 27 | 82 | 0.72 | 0.13 | 2.4 | 63 | 2.7 | 18 |

100 µm bands. In order to test this idea we obtained[5] the calibrated IRAS images of the AFGL 5142 region. The emission from the region is unresolved in all the four bands, but the peak of the point source seems to be slightly displaced in the four bands, being near IRS2 at 12 and 25 µm, and nearly coincident with the IRS1 position at 100 µm. Hence, all the information from the infrared data seem to tell us that the bright red source IRS2 is indeed a young object but not the most embedded and youngest of the region, while the faint NIR source IRS1 is probably one of the most embedded and youngest sources. The same conclusion has been obtained with our radio and molecular data which indicate that the H$_2$O masers, the UC radio continuum source, the peak of the high-density CS gas, and the center of the compact CO outflow all coincide with IRS1. It is clearly not possible to establish precisely the fraction of the FIR luminosity associated with each source, nevertheless we expect that a consistent fraction is indeed associated with IRS1. In the following we assume that the FIR luminosity of IRS1 is of the order of $10^3$ $L_\odot$. Combining this luminosity with the derived compact outflow parameters in Table 4, we obtain values for the outflow mechanical luminosity to stellar radiant luminosity ratio ($\geq 3 \times 10^{-3}$) and the outflow momentum flux to stellar radiant momentum flux ratio (45) (see eq. 19 and 20 of Garden et al. 1991). These values are consistent with the well-documented trends of increasing wind mechanical luminosity and wind force with increasing stellar bolometric luminosity (Lada 1985). This result suggests that central driving mechanism of the outflow in AFGL5142 is physically similar to outflows in other star-forming regions.

[5] The IRAS data were obtained using the IRAS data base server of the Space Research Organisation of the Netherlands (SRON) and the Dutch Expertise Center for Astronomical Data Processing funded by the Netherlands Organisation for Scientific Research (NWO). The IRAS data base server project was also partly funded through the Air Force Office of Scientific Research, grants AFOSR 86-0140 and AFOSR 89-0320.

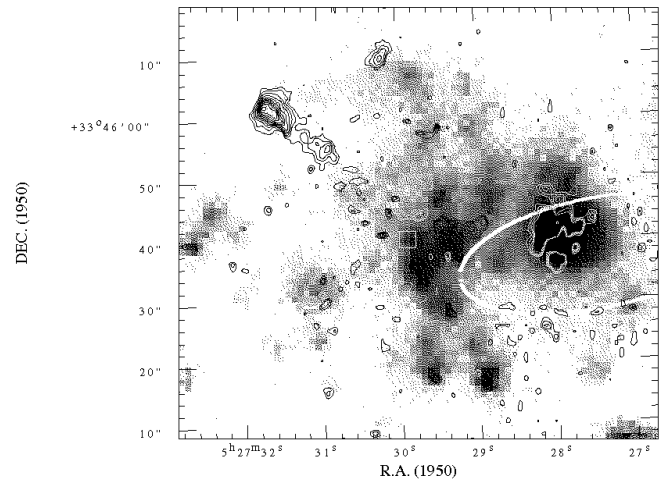

Fig. 10. AFGL 5142 central cluster. In greyscale is the K–band image, the contours are the H$_2$ emission, the ellipse the IRAS Point Source error box, the crosses represent the positions of the H$_2$O maser spots (the crosses are much larger than the position uncertainties) and the two concentric circles mark the peak of the radio continuum emission.

In Fig. 10 we present an overlay of the molecular hydrogen image (contours) on the K–band (greyscale). On the same figure we give also the error box of the IRAS Point Source (ellipse), and the position of the H$_2$O maser spots and that of the radio continuum peak (white square, see Fig. 2. The role of IRS1 as the most active source of the region is evident from the figure. Unfortunately from the J, H and K broad band data alone it is not possible to model the spectral energy distribution of the embedded source reliably and to determine the spectral type of the YSO, because the presence of the NIR excess prevents us from using the observed color indexes to derive the extinction of the central source. From the two J and H measurements alone, assuming that in these two bands the stellar photosphere emission extincted by the circum-stellar material dominates the measured flux, we would estimate a visual extinction greater than 15 magnitudes. Nevertheless it is likely that even at 1.25 $\mu$m and 1.65 $\mu$m the emission of the circumstellar material is important and this estimate of the extinction is very rough.

### 4.2. The association of the masers with the compact outflow

It is of interest to compare the energies (or the luminosities) involved in the different phenomena that occur in the surroundings of IRS1. As already explained it is difficult to quote a FIR luminosity for IRS1; a rough estimate of $\sim 10^3\ L_\odot$ can be used as indicative. The luminosity required to create the maser spots can be derived from the $J$-type shock collisional pumping model of Elitzur et al. (1989). Using their equation 4.5 the mechanical luminosity of the wind needed to create the conditions for 5 maser spots is of the order of 70 $L_\odot$. This value is rather uncertain since it depends on several model parameters (e.g. the size and aspect ratio of the maser) which are totally unknown, and is also independent of distance. However, it should be correct as an order of magnitude. A mechanical wind luminosity between 5 to 10% of the total FIR luminosity tells us that a considerable fraction of the stellar energy must go into the shock compression to create the masers. Most of this energy will be reprocessed into FIR continuum by heated dust. In fact, the mechanical luminosity of the outflow is only 2.7 $L_\odot$. This means that only 4% of the energy needed to produce the maser goes into the acceleration of the gas in the molecular outflow lobes, with a rather low efficiency, while the rest is dissipated into heat. The detection of a high excitation molecule such as CS $J$=7→6 with $T_{mb}$ = 8 K may suggest high kinetic temperature in the inner core, as also indicated by the high temperature found in the NH$_3$ line (Estalella et al. 1993). From the H$_2$ fluxes of Table 3 the integrated luminosity of the H$_2$ jets in the S(1) 1 → 0 line is $\sim$0.02 $L_\odot$. Interestingly, this value is two orders of magnitude smaller than the mechanical luminosity of the CO outflow; since the jet is a factor of 5 smaller in size, the outflow and the jet contain approximately equal mechanical luminosity per unit volume. Finally, the luminosity of all the H$_2$O masers is 5.7 $10^{-6}\ L_\odot$. The ratio of this value with the total FIR luminosity is $\sim 5\ 10^{-9}$, in very good agreement with the average value found by Felli et al. (1992) from a large sample. However, the present analysis points out that only a small fraction of the total FIR luminosity is directly related to the creation of the masers and of the H$_2$ jets-molecular outflow. The correlation found between maser luminosity and total FIR luminosity over a large range of luminosities suggests that this fraction should not vary appreciably with the luminosity. The ratio of H$_2$O maser luminosity to mechanical luminosity of the CO outflow is $\sim 2.1\ 10^{-6}$, nearly identical to the average value found by Felli et al. (1992) for a large sample of regions.

Although there is no simple velocity pattern observed across the more distant maser components (C1, C2 and C5), they lie along an axis that is roughly parallel to the jet observed in H$_2$ and to the compact bipolar molecular outflow. The proper motions suggested for C2 and C5 also lie outward along this axis. To summarize, the physical picture that emerges from these observations, is that the ionized stellar wind (detected in the 8.4 GHz continuum) ultimately generates shocks in the gas surrounding the YSO (IRS1) where the density is high enough (as traced by the CS $J$=7→6 transition) to collisionally-pump the H$_2$O maser line, while at the same time drives the highly-collimated H$_2$ jet and the compact bipolar molecular outflow seen at larger distances. The picture might be different for C3 and C4, the ones within a few hundred AU of the YSO, where we may see the interaction of the accreting disk material with the stellar wind.

## 5. Conclusions

Our new observations provide a clearer picture of the star forming region AFGL 5142 and a better comprehension of the effects generated by the interaction of the energy sources (the YSOs) with the surrounding molecular environment (i.e. the water masers, the regions of shocked H$_2$ and the CO molecular outflows).

A cluster of IR sources is found in the center of the region, with a radius of 0.3 pc and with a K-band source density about 10 times that over the rest of the observed field. Many of the sources in the cluster show a strong IR excess, typical of YSOs still surrounded by their parent dust/molecular cloud. Two of the cluster members, separated by about 30", have received more attention: IRS1, which has the strongest IR excess and IRS2, the brightest member at K band (2.7 magnitudes more luminous than IRS1). IRS2 seems to be at the center of an extended and well collimated CO outflow. It also coincides with an IRAS point source. Several pieces of evidence suggest that IRS2 may represent a YSO in an advanced phase of evolution: there are no traces of UC H II regions or of maser emission in IRS2, an HH object has been found in its proximity, and it lies away from the peak of the high density molecular tracers. All these facts imply that the region of the molecular cloud from which IRS2 formed has been dispersed, the extinction is smaller and that the only remnant of the star formation process is the large scale collimated outflow.

In contrast, IRS1 possesses all the characteristic aspects of a YSO in an early evolutionary phase. It is at the center of a compact CO bipolar outflow, it coincides with an almost unresolved thermal continuum source (most probably an ionized wind), it is at the center of a cluster of H$_2$O masers (two of which are within a few hundred AU and two others with indication of proper motion in the outward direction), it coincides with a small and dense molecular cloud core and it is at the origin of H$_2$ jets. The H$_2$O maser emission occurs only in the immediate surroundings of the YSO, where a substantial fraction of the YSO energy required for their formation is available. This tight spatial connection between masers and YSOs is on the same lines of that found by Testi et al. (1994) in a large sample of $H_2O$ masers. The maser velocities are always much lower than those of the two outflow lobes, indicating that they occur in the region where the acceleration begins. No maser formation occurs either in the region where the $H_2$ is shock excited, or at the outer edges of the outflow where the high velocity molecular gas hits the surrounding material, most probably because there is no longer sufficient energy to create $J$-type shocks.

Higher resolution, single dish and interferometric molecular line observations will be important to explore this dense molecular core as well as to better separate the two bipolar outflows. Furthermore, additional high-resolution $H_2O$ maser observations and single dish studies are needed to confirm the proper motions implied by the two epochs analyzed here and to study the dependence of variability on the distance of the maser from the YSO.

*Acknowledgements.* We thank the CSO staff for their ongoing assistance during this observing project and offer our deep though posthumous appreciation to Larry Strom for his assistance at the summit over the past few years. GBT acknowledges support from NSF under grant AST-9117100.